# On The Development of Low-power MAC Protocol for WBANs

Sana Ullah *Member* IAENG, Pervez Khan, and Kyung Sup Kwak

*Abstract-* **Current advances in wireless communication, microelectronics, semiconductor technologies, and intelligent sensors have contributed to the development of unobtrusive WBANs. These networks provide long term health monitoring of patients without any constraint in their normal activities. Traditional MAC protocols do not accommodate the assorted WBAN traffic requirements in a power efficient manner. In this paper, we present a brief discussion on the development process of a low power MAC protocol for WBANs. We observe the behavior of a beacon-enabled IEEE 802.15.4 for on-body sensor networks. We further propose a low power technique called *traffic based wakeup mechanism* for a WBAN that exploits the traffic patterns of the BAN Nodes to ensure power efficient and reliable communication.**

## I. Introduction

Cardiovascular disease is the foremost cause of death in the United States and Europe since 1900. More than ten million people are affected in Europe, one million in the US, and twenty two million people in the world [1] – [3]. The number is projected to be triple by 2020, or 20% of the Gross Domestic Product (GDP). The ratio is 17% in South Korea and 39% in UK [4]-[5]. The healthcare expenditure in the US is expected to increase from 2.9 trillion in 2009 to 4 trillion in 2015 [6]. The impending health crisis attracts researchers, industrialists, and economists towards optimal and quick health solutions. The non-intrusive and ambulatory health monitoring of patient's vital signs with real time updates of medical records via internet provides economical solutions to the health care systems.

A WBAN contains a number of portable, miniaturised, and autonomous sensor nodes that monitors the body function for sporting, health, entertainment, and emergency applications. It provides long term health monitoring of patients under natural physiological states without constraining their normal activities. In-body sensor networks allow communication between implanted devices and remote monitoring equipments. They may be used to collect information from Implantable Cardioverter Defibrillators (ICDs) in order to detect and treat ventricular tachyarrhythmia[1] and prevent Sudden Cardiac Death (SCD) [7].

A number of ongoing projects such as CodeBlue, MobiHealth, and iSIM have contributed to establish a proactive and unobtrusive WBAN system [8]-[10]. A system architecture presented in [11] performs real time analysis of sensor's data, provides real time feedback to the user, and forwards the user's information to a telemedicine server. UbiMon aims to develop a smart and affordable health care system [12]. MIT Media Lab is developing MIThril that gives a complete insight of human-machine interface [13] HIT lab focuses on quality interfaces and innovative wearable computers [14]. IEEE 802.15.6 aims to provide low-power in-body and on-body wireless communication standards for medical and non-medical applications [15]. NASA is developing a wearable physiological monitoring system for astronauts called LifeGuard system [16]. ETRI focuses on the development of a low power MAC protocol for a WBAN [17].

In this paper, we focus on the development of a low power MAC protocol for in-body and on-body sensor networks. We use the terms BAN Node (BN) and BAN Network Coordinator (BNC) for the sensor node and the network coordinator in a WBAN. The rest of the paper is organized into five sections. Section II presents the WBAN traffic classification. Section III and IV discuss MAC for on-body and in-body sensor networks including a performance analysis of several low power MAC protocols. Section V presents a low power technique called *traffic based wakeup mechanism*. Section VI concludes our work.

## II. Traffic Classification

The assorted WBAN traffic requires sophisticated and power efficient techniques to ensure safe and reliable operation. Existing MAC protocols such as SMAC [18], TMAC [19], IEEE 802.15.4 [20], and WiseMAC [21] give limited answers to the heterogeneous traffic. The in-body BNs do not appreciate synchronized wakeup periods because they confine the accommodation of sporadic emergency events. Medical data usually needs high priority and reliability than non-medical data. In case of emergency events, the BNs should be able to access the channel in less than one second [22]. IEEE 802.15.4 Guaranteed Time Slots (GTS) can be utilized to handle time critical events but they expire in case of a low traffic. Furthermore, some in-body BNs have high data transmission frequency than others. We classify the entire WBAN traffic into Normal, On-demand,

Manuscript received January 4, 2009. This work was supported by the IT R&D program of Ministry of Knowledge Economy (MKE)/Institute of Information Technology Assessment (IITA), [2008-F-050-01, Development of WBAN system for In-body and On-body Sensor Networks]
Sana Ullah, Pervez Khan, and Kyung Sup Kwak are with Graduate School of IT and Telecommunications, Inha University, 253 Yonghyun-Dong, Nam-Gu, Incheon 402-751, South Korea (sanajcs@hotmail.com, pervaizkanju@hotmail.com, and kskwak@inha.ac.kr)

[1] Ventricular tachyarrhythmia are abnormal patterns of electrical activity originating within ventricular tissue.

and Emergency traffics as given in Fig 1. The normal traffic is further classified into High, Medium, and Low traffics.

*A- Normal Traffic:* Normal traffic is the data traffic in a normal condition with no time critical and on-demand events. This includes unobtrusive and routine health monitoring of a patient for diagnosis and treatment of many diseases such as gastrointestinal tract, neurological disorders, cancer detection, handicap rehabilitation, and the most threatening heart disease. Some BNs have frequent wakeup periods and thus are designated as high traffic BNs. For example, an ECG node may send data 4 times per hour, while other BNs may send 4 times a day. The ECG node is thus designated as a high traffic BN. However, the normal traffic classification from high to low and vice-versa depends on the application requirements. The normal data is collected and processed by the BNC.

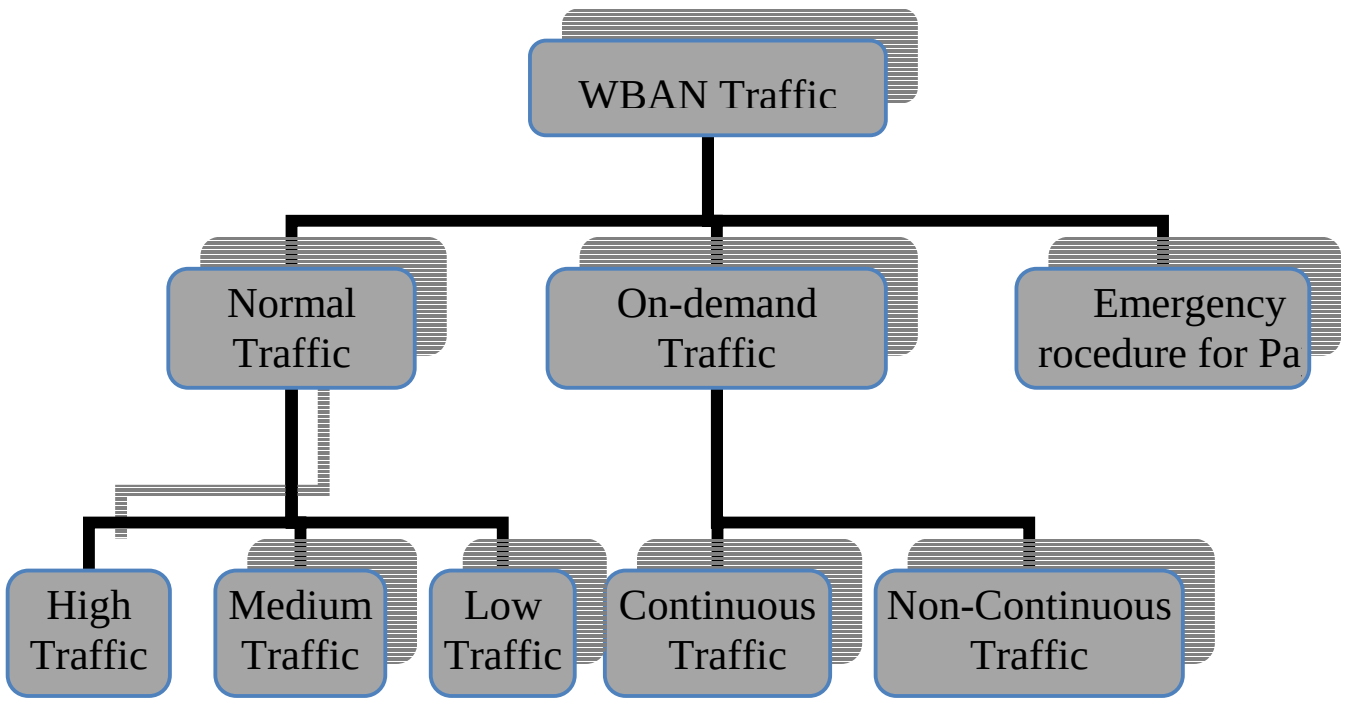

*Fig 1. WBAN Traffic Classification*

*B- On-demand Traffic:* On-demand traffic is initiated by the BNC or doctor to know certain information, mostly for the purpose of diagnosis and prescription. This is further divided into continuous (in case of surgical events) and non-continuous (when limited information is required).

*C- Emergency Traffic:* This is initiated by BNs when they exceed a predefined threshold and should be accommodated in less than one second. This kind of traffic is not generated on regular intervals.

## III. ON-BODY SENSOR NETWORK

On-body sensor networks comprise of miniaturized and non-invasive sensor nodes that are used for various applications, ranging from medical to interactive gaming and entertainment applications. They use Wireless Medical Telemetry Services (WMTS), unlicensed ISM, and UWB bands for data transmission. WMTS is a licensed band designated for medical telemetry system. Federal Communication Commission (FCC) urges the use of WMTS for medical applications due to fewer interfering sources. However, only authorized users such as physicians and trained technicians are eligible to use this band. Further the restricted WMTS (14 MHz) bandwidth cannot support video and voice transmission. The alternative spectrum for medical applications is to use 2.4 GHz ISM band that includes guard bands to protect adjacent channel interference. UWB is a promising candidate to satisfy the baseline power consumption requirements of on-body sensor networks.

The design and implementation of a low power MAC protocol for on-body sensor networks have been a hot research topic for the last few years. A novel TDMA protocol for on-body sensor network called H-MAC exploits the biosignal features to perform TDMA synchronization and improves the energy efficiency [23]. Other protocols like WASP, CICADA, and BSN-MAC are investigated in [24] – [26]. The performance analysis of a non-beacon IEEE 802.15.4 is adapted to extend lifetime of a node from 10 to 15 years [27]. This work considers low upload/download rates, mostly per hour. Further the data transmission is based on periodic intervals, which limits the performance to certain applications. There is no reliable support for on-demand and emergency traffics. Intel Corporation has carried out a series of experiments to analyze the performance of IEEE 802.15.4 for an on-body sensor network [28]. They deployed a number of Intel Mote 2 [29] devices on chest, waist, and the right ankle. Experimental results show that the packet delivery ratios are 100%, 99%, and 84% at waist, chest, and ankle respectively. The connection between ankle and waist cannot be established, even for a short distance of 1.5 meters.

We present the performance analysis of a beacon-enabled IEEE 802.15.4, Preamble-Based TDMA [30], and SMAC protocols for on-body sensor networks. Our analysis is verified by extensive simulations using NS-2. Simulation results show that IEEE 802.15.4, when configured in a beacon-enabled mode, outperforms SMAC and PB-TDMA as shown in Fig 2. However, the precise location of BNs and the body position influence the packet delivery ratio.

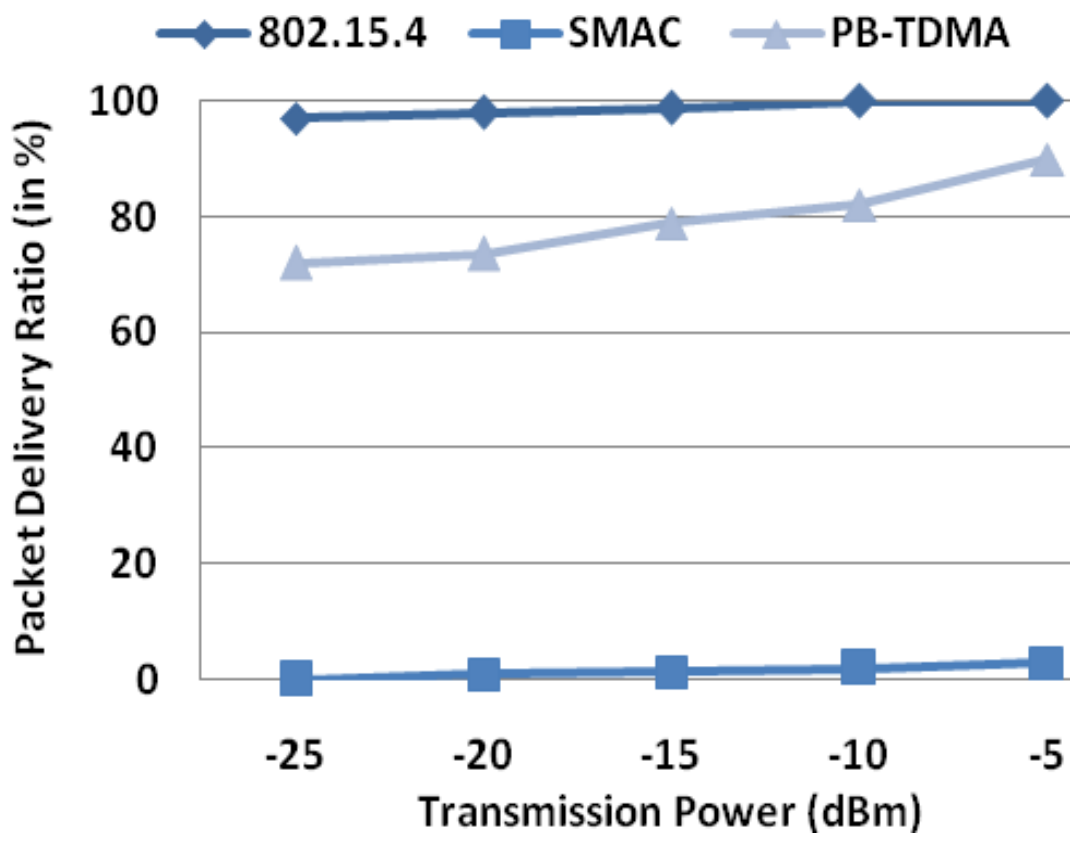

*Fig 2. Packet Delivery Ratio of IEEE 802.15.4, SMAC, and PB-TDMA for On-body Sensor Networks*

## IV. IN-BODY SENSOR NETWORK

In-body sensor networks allow communication between implanted BNs and external monitoring equipments. Unlike on-body BNs, in-body BNs are implanted under human skin where the electrical properties of the body affect the signal propagations. The human body is a medium that poses many wireless transmission challenges. The body is composed of several components that are not predictable and will change. Monitoring in-body functions and the ability to communicate with implanted therapeutic devices, such as pacemakers, are essential for their best use. Applications include monitoring and programme changes for pacemakers and ICDs, control of bladder function, and restoration of limb movement. The space within a body is very limited and the available materials are few. These applications may require continuous or occasional one or two-way transmission. Some applications will require a battery where the current drain must be low so as not to reduce the working life of the implant function.

Zarlink semiconductor has introduced a wireless chip that supports a very high data rate RF link for communication with an implantable BN [31]. The ZL70101 ultra-low power transceiver chip satisfies the power and size requirements for implanted communication systems and operates in 402-405 MHz MICS band [32]-[33]. Other frequency bands such as 916MHz, 1.5GHz, and Ultra-Wideband are also considered for in-body sensor networks [34]-[36]. The use of open air models for implant communication is discouraged in [37]. A Finite-Difference Time-Domain (FDTD) is used to calculate the power deposition in a human head and is measured by the SAR in W/Kg [38]. However, the distance from the implant source has not been discussed. For 0.05% duty cycle at 400 kbps, the SubQore radio architecture from Cambridge university consumes a peak current less than 1.7mA [39]. An implantable medical microsystems for interfaces to the central nervous system is presented in [40].

The diverse nature of in-body BNs together with the electrical properties of the human body influences the development process of a low power and heterogeneous MAC for in-body sensor networks. The data rate of implanted BNs varies, ranging from few kbps in pacemaker to several Mbps in capsular endoscope. Fig 3 explains the in-body sensor networks and implementation issues [41].

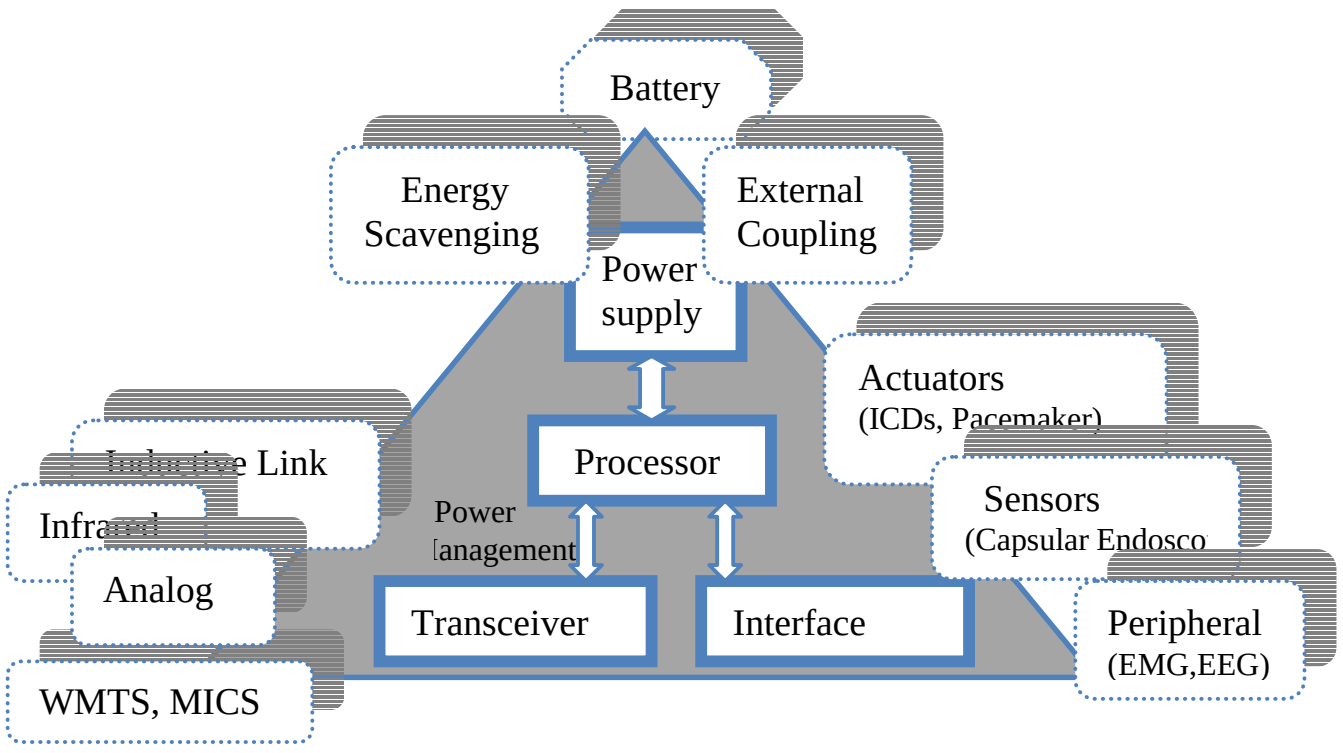

*Fig 3. Implementation issues in the in-body sensor network*

In the in-body sensor network, critical data requires low latency and high reliability. The solution is to adjust initial back-off windows for critical and non-critical traffics [42]. Non-critical BNs have larger initial back-off window than critical BNs, i.e., ($W_0^{\alpha} \leq W_0^{\beta}$) where $W_0^{\alpha}$ is initial back-off window for critical BNs $\alpha$ and $W_0^{\beta}$ is initial back-off window for non-critical BNs $\beta$. Fig 4 illustrates the average packet latency for a number of in-body BNs. The smaller initial back-off window for the critical BNs results in lower latency. The Fig also shows packet latency of a standard CSMA/CA scheme used in IEEE 802.15.4.

The traditional CSMA/CA technique does not provide reliable solution in multiple piconets [41]. For a threshold of -85dBm and -95dBm, the on-body BNs cannot see the activity of in-body BNs when they are away at 3 meters distance from the body surface as given in Fig 5. However, in 3 meters or less distance, the CCA works correctly in the same piconet. The main reason of improper energy detection is that the path loss inside human body is much higher than in free space. However, the performance of other CCA techniques, i.e., Preamble Detector and De-correlation Based Detector can be investigated for in-body BNs. Preamble detector indicates a busy channel only on detecting the preamble part of the frame. De-correlation on the other hand, looks for certain spreading characteristics other than preamble.

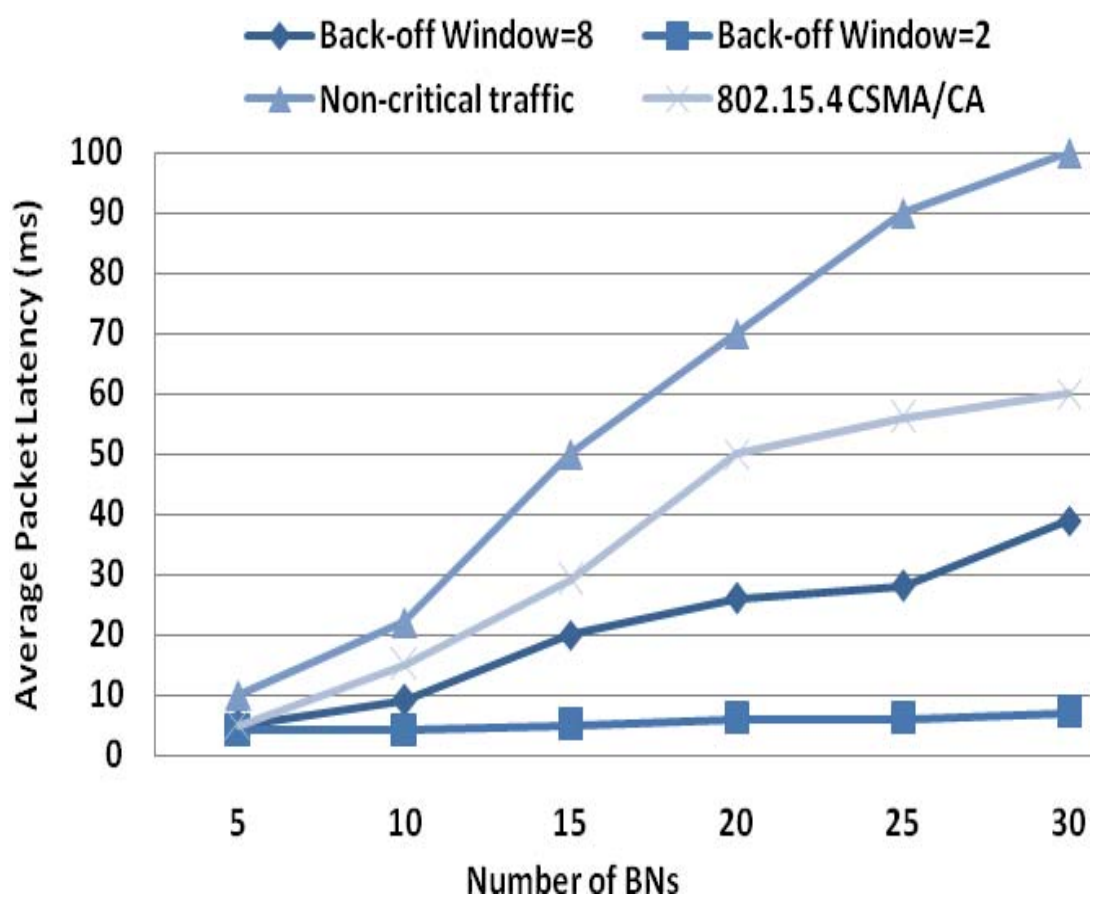

*Fig 4. Average Packet Latency for Critical and Non-critical BNs*

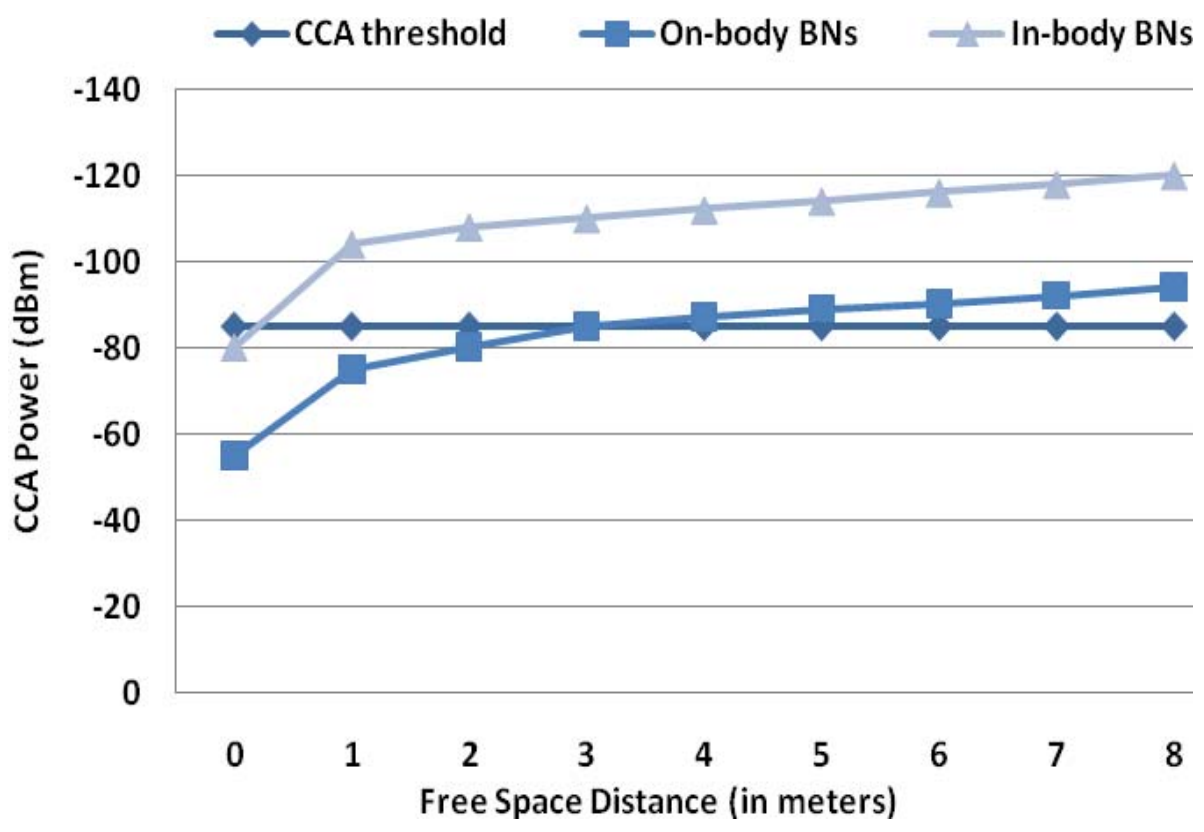

*Fig 5. CCA for In-body and On-body BNs*

## V. TRAFFIC BASED WAKEUP MECHANISM FOR WBAN

The heterogeneous BNs require power efficient and dynamic wakeup techniques for reliable operation. We propose a low power technique called *traffic based wakeup mechanism* that exploits the traffic patterns of BNs to accommodate the assorted WBAN traffic. The initial wakeup patterns are either predefined (by the company) or created and modified (by the BNC). Table 1 shows the aforementioned traffic classification and the corresponding solutions.

*Table 1. Solutions to the Classified WBAN Traffic*

| **Traffic** / **Device** | **Normal Traffic** | | | **On-demand Traffic** | **Emergency Traffic** |
|---|---|---|---|---|---|
| | High | Medium | Low | | |
| **BAN Nodes (BNs)** | Send data based on the **Traffic-based Wake-up Table** | | | Receives a **Wakeup Radio Signal** from the BNC and respond | Send a **Wake-up Radio Signal** to the BNC in case of emergency |
| **BAN Network Controller (BNC)** | Send data based on the **Traffic-based Wake-up Table** | | | Send a **Wake-up Radio Signal** to BNs | Receives a **Wake-up Radio Signal** and respond |

The wakeup patterns of all BNs are organized into a table called *traffic based wakeup table*. The table is maintained and modified by the BNC according to the requirements. Based on the BNs wakeup patterns, the BNC can also calculate its own wakeup pattern. This could save significant energy at the BNC. The BNC does not need to stay active when there is no traffic from the BNs. The designation of normal traffic levels, i.e, high, medium, and low traffics BNs depend on the application. For example, some BNs may repeat their wakeup patterns per BAN superframe and thus are designated as high traffic BNs. Others may repeat per hour (medium traffic BNs) and per day (low traffic BNs). Fig 6 shows the repetition of wakeup patterns in a multiple number of BAN superframes . In this case, BN-1 is a high traffic node having a wakeup pattern equal to *10x* where *x* is the duration of a single BAN superframe. BN-3 is the low traffic BN having a wakeup pattern equal to *43x*. BNs having the same wakeup patterns contend for the channel.

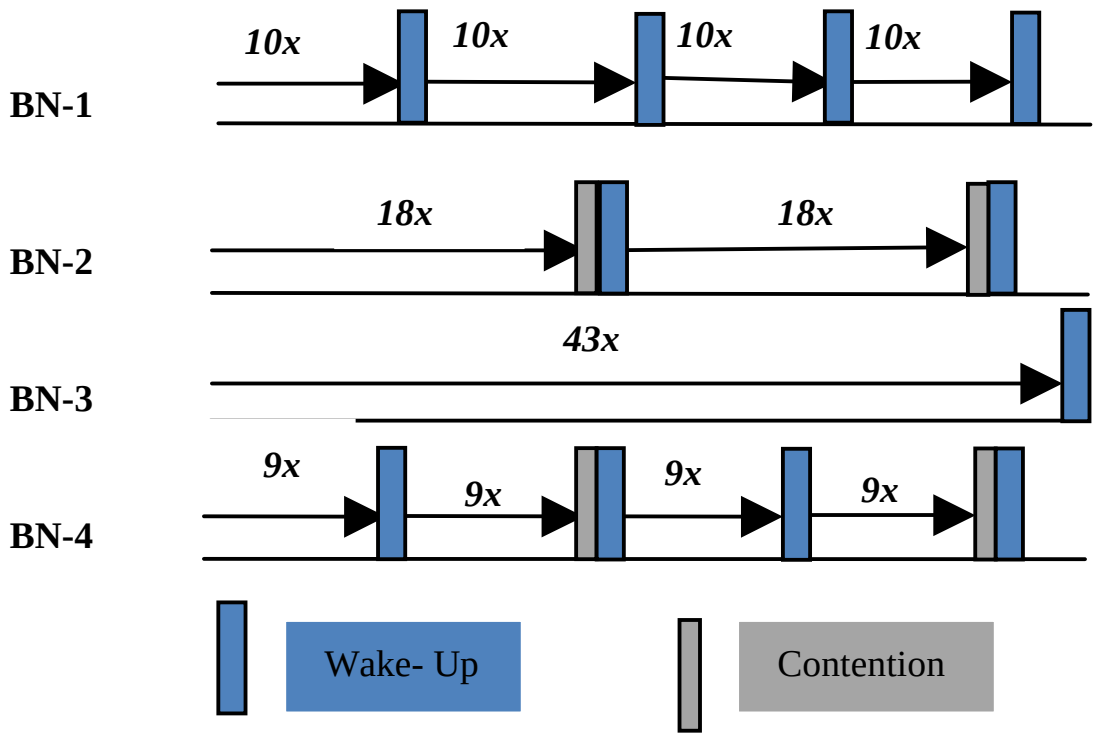

Fig 6. Superframe Based Wakeup Patterns

In case of TDMA, the wakeup patterns represent the corresponding slots in the TDMA superframe. The data slots are active according to the wakeup patterns. For example, *10x* would allow using the BN-1 data slot after 10 TDMA superframes.

For emergency and on-demand traffics, the BNs and the BNC send a wake-up signal for a very short duration to each other. However, traditional wake-up radio concepts have several limitations when considered in the in-body sensor networks. They are not able to wake-up a particular node. All BNs wake-up in response to a single wake-up signal, which is not the required environment. The use of different radio frequencies to wake-up a particular node may provide an optimal solution.

## VI. CONCLUSIONS

In this paper, we presented a technical discussion on low power MAC protocols for a heterogeneous in-body and on-body sensor networks. The traffic is classified into normal, on-demand, and emergency traffics. For emergency events, the BNs should be able to access the channel in less than one second. Furthermore, the beacon enabled IEEE 802.15.4 transcended other low power MAC protocols in terms of throughput. However, it is not sufficient to accommodate the entire WBAN traffic characteristics including time critical events. Time critical events require low latency and high reliability. The adjustment of initial back-off window in the traditional CSMA/CA technique provides limited solution. But the use of energy detection CCA for in-body sensor networks is not a reliable approach due to high path loss inside the human body. We proposed a low power *traffic based wakeup mechanism* for a WBAN that exploits the traffic patterns of BNs to accommodate the entire traffic classification. In future, we will extend our technique towards a complete implementation. Our proposed approach will provide a power efficient and innovative solution for in-body and on-body sensor networks.